\documentclass[11pt]{article}
\usepackage{graphicx,amssymb,amsmath}

\newcommand{\halfrightsect}[2]{[#1,#2)}

\def\idrm#1{\ensuremath{\mathrm{#1}}}

\def\etal{\emph{et~al.}}
\def\ie{\emph{i.e.}}

\newenvironment{itemize*}%
  {\vspace{-1ex}\begin{itemize}%
    \setlength{\itemsep}{-1ex}%
    \setlength{\parsep}{0pt}}%
  {\end{itemize}}

\newcommand{\col}{\idrm{col}}

\newcommand{\cT}{{\cal T}}
\newcommand{\cC}{{\cal C}}

\newtheorem{theorem}{Theorem}
\newtheorem{obs}{Observation}

\begin{document}
\title{Searching for Frequent Colors in Rectangles}
\author{Marek Karpinski\thanks{Dept. of Computer Science, University of Bonn.  Email {\tt marek@cs.uni-bonn.de}.} 
\and 
Yakov Nekrich\thanks{Dept. of Computer Science, University of Bonn. Email {\tt yasha@cs.uni-bonn.de}.}
}
\date{}
\maketitle

\begin{abstract}
We  study a new variant of colored orthogonal range searching
 problem: given a query rectangle $Q$ all colors $c$, such that at least a 
fraction $\tau$ of all points in $Q$ are of color $c$, must be reported. 
We describe several data structures for that problem that use pseudo-linear
 space and answer queries in poly-logarithmic time.
\end{abstract}

\section{Introduction}
The colored range reporting problem is a variant of the  
range searching problem in which every point $p\in P$ 
is assigned a color $c\in C$. The set of points $P$ is pre-processed 
in the data structure so that for any given rectangle $Q$ all distinct 
\emph{colors} of points in $Q$ can be reported efficiently. 
In this paper we consider  a variant of this extensively studied problem 
in which only frequently occurring colors must be reported.

We say that a color $c\in C$ \emph{$\tau$-dominates} rectangle $Q$ if at 
least a $\tau$-fraction of points in $Q$ are of that color:
$|\{\,p\in P\cap Q\mid \col(p)=c\,\}| \leq \tau|P\cap Q|$,
where $\col(p)$ denotes the color of point $p$.  
We consider several data structures that allow us to report 
colors 
 that dominate $Q$ 
\footnote{Further we will assume that parameter $\tau$ is fixed and  simply 
say that a color $c$ dominates rectangle $Q$ .}.  
\\
{\bf Motivation}
Standard colored range reporting problem arises in many applications. 
Consider a database in which every object is characterized by 
several numerical values (point coordinates) and some attribute (color). 
For instance the company database  may contain information about 
age and salary of each employee. The attribute associated with each 
employee is her position. The query consists in reporting 
all different job types for all employees with salary between 
40.000 and 60.000 who are older than 40 and younger than 60 years old.
Colored range reporting  also  occurs naturally  in 
 computational biology applications:  each amino acid is associated with 
certain attributes (hydrophobic, charged, etc.). We may want to report 
different attributes associated with amino acids in certain 
range~\cite{MGM05}. 

However, in certain applications we are not interested in all attributes 
that occur in the query range. Instead, we may be interested in 
reporting the \emph{typical} attributes. For instance, in the first 
example above we may wish to know all job types, such that at least 
a fraction $\tau$ of all employees with a given salary and age range 
have a job of this type. In this paper we describe data structures that 
support such and similar queries.\\
{\bf Related Work.} 
Traditional colored range reported queries can be efficiently 
answered in one, two, and three dimensions. 
There are data structures that use pseudo-linear space 
and answer one- and  two-dimensional colored range 
reporting queries in $O(\log n + k)$ time~\cite{GJS95},~\cite{JL93}
 and   three-dimensional 
colored  queries in $O(\log^2 n + k)$ time~\cite{GJS95},  
where $k$ is the number of colors. 
A semi-dynamic  data structure of Gupta~\etal~\cite{GJS95} supports 
two-dimensional queries in $O(\log^2 n +k)$ time and 
insertions in $O(\log^3 n)$ amortized time.
Colored orthogonal range reporting queries in $d$ dimensions can be answered 
in $O(\log n +k)$ time with a data structure that uses 
$O((n^{1+\varepsilon}))$ space~\cite{AGM02}, but no efficient 
 pseudo-linear space data structure is known for $d>3$.

De Berg and Haverkort~\cite{BH03} consider a variant of the 
colored range searching in which only \emph{significant} colors must 
be reported. A color $c$ is \emph{significant} in rectangle $Q$ if 
 at least a fraction $\tau$ of points of that color 
belong to  $Q$, $|\{\,p\in Q\cap P\mid\col(p)=c\,\}| \geq \tau|\{p\in P\mid 
\col(p)=c\,\}|$.  For $d=1$, de Berg and Haverkort~\cite{BH03}
 describe a linear space data structure 
that answers queries in $O(\log n +k)$ time, where $k$ is the number of 
signficant colors. For $d\geq 2$ signficant queries can be answered
 approximately: in $O(\log n + k)$ time we can report all a set of 
colors such that each color is $(1-\varepsilon)\tau$-significant for 
a fixed constant $\varepsilon$ and all $\tau$-significant colors 
are reported. The only known  data structure that efficiently answers 
exact significance queries uses cubic space~\cite{BH03}.  
\\
The problem of finding the elements occurring  at least $\tau n$ times 
in a \emph{stream} of data was studied in the context of 
streaming algorithms~\cite{MG85},~\cite{DLM02},~\cite{KSP03}. 
It is possible to find all elements that occur at least $\tau n$ times 
in the multi-set of $n$ elements with an algorithm that $O(1/\tau)$ space 
and with two passes through the data~\cite{MG85},~\cite{DLM02},~\cite{KSP03}. 
However, any algorithm that performs only one pass through the data 
must use  $\Omega(m\log\frac{n}{m})$ bits of space, 
where $m$ is the number of different elements (colors in our terminology) 
in the multi-set~\cite{KSP03}.  
\\
{\bf Our Results}
In this paper we show that we can find domination colors 
 in an arbitrary $d$-dimensional rectangle 
in poly-logarithmic time using a pseudo-linear space data structure. 
\begin{itemize*}
\item
We describe a static $O(\tau n)$ space data structure that supports 
one-dimensional queries in $O(\tau \log n \log \log n)$ time.
A static $O(\tau n\log \log  n)$ space data structure supports 
one-dimensional domination  queries in $O(\tau \log n)$ time.
\item
In the case when all coordinates are integers bounded by $U$, 
there is a $O(\tau n)$ space static data structure that supports 
one-dimensional domination queries 
in $O(\tau \log  \log n \log \log U)$ time
\item
There is a dynamic $O(\tau n)$ space data structure that supports 
one-dimensional domination queries 
 and insertions in $O(\tau \log n)$ time  and deletions in 
$O(\tau \log n)$ amortized time.  We can reduce the update 
time to (amortized)  $O(\log n)$ by increasing the space usage 
to $O(\tau n\log n)$
\item
There is a data structure that supports domination queries 
in $d$ dimensions in $O(\tau \log^{d} n )$ time 
and uses $O(\tau n\log^{d-1} n)$ space 
\item
There is a dynamic data structure that answers 
domination queries in $d$ dimensions in $O(\tau \log^{d+1} n)$ 
time, uses  $O(\tau n\log^{d-1} n)$ space, and supports 
insertions in $O( \tau \log^{d+1} n)$ time and deletions in
 $O(\tau \log^{d+1} n )$ amortized time
\end{itemize*}
We describe static and dynamic data structures for one-dimensional 
domination queries in sections \ref{sec:static} and \ref{sec:dynamic}.
Data structures for multi-dimensional domination queries are described in 
section~\ref{sec:multidim}.
\section{Static Domination Queries in One Dimension }
\label{sec:static}
The following simple property plays an important role in all data structures 
for domination queries.
\begin{obs}\label{obs:decompose}
If $Q=Q_1\cup Q_2$, $Q_1\cap Q_2=\emptyset$,  and color $c$ is dominant 
in $Q$, then either $c$ is dominant in $Q_1$ or $c$ is dominant in $Q_2$.
\end{obs}
Due to this property  a query  on a set $Q$ 
can be \emph{decomposed} into queries  on some disjoint 
sets $Q_1,\ldots, Q_p$ such that $\cup Q_i=Q$ and $p$ is a constant:
we find the dominating colors for each $Q_i$ and for each 
color $c$ that dominates some $Q_i$ we determine 
whether $c$  dominates $Q$ by a range counting query. 



Our data structure is based on the same approach as exponential search 
trees~\cite{AT07}.
Let $P$ be the set of all points. In one-dimensional case we do not distinguish between a point and its coordinate. 
$P$ is divided into $\beta_n$ intervals $I_1,\ldots, I_{\beta_n}$ so that 
each $P_i=P\cap I_i$ contains between $n^{2/3}/2$ and $2n^{2/3}$ points and 
$\beta_n=\Theta(n^{1/3})$.
Let $l_i$ and $r_i$ denote the left and right bounds of interval $I_i$. For each $1\leq i\leq j\leq\beta$, the list $L_{ij}$ 
contains the set of colors that dominate $[l_i,r_j]$. 
We denote by $n_{ij}$ the total number of points in $[l_i,r_j]$. 

Each interval $I_i$ is recursively subdivided in the same manner: an interval 
that contains $m$ points is divided into $\beta_m$ subintervals and 
each subinterval contains between $m^{2/3}/2$ and $2m^{2/3}$ points.
If some interval $I_j$ is divided into $I_{j,1},\ldots, I_{j,\beta}$, then 
we say that $I_j$ is a parent of $I_{j,i}$ ($I_{j,i}$ is a child of 
$I_j$). The tree $\cT$ reflects the division of  intervals 
into sub-intervals: each tree node $u$ corresponds to an interval $I_u$ 
and a node $u$ is a child of $v$ if and only if $I_u$ is a child of $I_v$. 
The root of $\cT$ corresponds to $P$ and leaves of $\cT$ correspond 
to points of $P$. 
The node of depth $i$ contains $n^{(2/3)^i}$ points. Hence, the node 
of depth $\log_{\frac{3}{2}}\log n$  contains $O(1)$ points and
the height of $\cT$ is $O(\log \log n)$.
For every color $c$, we also store all points of color $c$ in a data 
structure that supports range counting queries. 

Consider a query $Q=[a,b]$. Let $l_a$ and $l_b$ be the leaves of $\cT$  
in which $a$ and $b$ are stored, and let $q$ be the lowest common 
ancestor of $l_a$ and $l_b$. 
The search procedure visits all nodes on the path from $l_a$ to $q$ 
($l_b$ to $q$); for each visited node $u$ we construct the set 
of colors $S_u$, such that every $c\in S_u$ dominates $I_u \cap [a,b]$. 
We also compute  the total number of points in $I_u \cap [a,b]$.
Let $u$ be the currently visited node of $\cT$ situated between 
$l_b$ and $q$,  and suppose that 
the node $v$ visited immediately before $u$ is the $(i+1)$-st child of $u$.
Due to Observation~\ref{obs:decompose} only colors stored in $L_{1i}$ 
and $S_v$ may dominate $I_u \cap Q$.
For each color $c$ in $L_{1i}\cup S_v$ we count how many 
times it occurs in $I_u \cap Q$ using the range counting data structure 
for that color.
Thus we can construct $S_u$ by answering 
at most $2\tau$ counting queries. 
Nodes between $l_a$ and $q$ are processed in the same way.
Finally, we examine all colors in sets $S_p$ and $S_r$ and 
list $L_{ij}$ of the node $q$, where $p$ and $r$ are nodes on the paths 
from $q$ to $l_a$ and $l_b$ respectively, $p$ is the 
$i$-th child of $q$, and $r$ is the $j$-th child of $q$. 
The search procedure visits $O(\log \log n)$ nodes 
and answers $O(\tau \log \log n)$ counting queries.  
Hence, queries can be answered in $O(\log n \log \log n)$ time. 

If an interval $I$ contains $m$ points, then all lists $L_{ij}$ contain  
$O(m^{2/3})$ elements.  Data structures for range counting queries 
use $O(n)$ space. Therefore the space usage of our data structure 
is $O(n)$. 

We can reduce the query time to $O(\log n)$ by storing range counting 
data structures for each interval: for every interval $I_u$ and every 
color $c$, such that $\{\, p\in P\cap I_u \mid\col(p)=c\,\}\not=\emptyset$,
we store a data structure that supports range counting queries in 
time $O(\log |I_u|)$. The total number of colors in all intervals 
$I_u$ for all nodes $u$ situated on the same level of tree $\cT$ 
does not exceed the number of points in $P$. Therefore the total 
number of elements in all range counting data structures is
 $O(n\log \log n)$.  The query is processed in the same way as described
 above. We must answer $O(\tau)$ counting queries on $I_q$, 
$O(\tau)$ range counting queries on children of $I_q$, 
$O(\tau)$ range counting queries on children of children of $I_q$, etc. 
Therefore the query time is $O(\tau(\log(|I_q|) + \log(|I_q|^{2/3}) +\log(|I_q|^{4/9})+
\ldots))=O(\tau\sum (2/3)^i \log n)=O(\tau\log n)$. 

We obtain the following result
\begin{theorem}\label{theor:static1d}
There exists a $O(\tau n \log \log n)$ space data structure  that 
supports  one-dimensional domination  queries in $O(\tau \log n)$ 
time. There exists a $O(\tau n)$ space data structure  that 
supports one-dimensional  domination  queries in $O(\tau \log n \log \log n)$ 
time. 
\end{theorem}

In the case when all point coordinates are integers bounded 
by a parameter $U$ we can easily answer one-dimensional counting queries 
in $O(\log \log U)$ time. As shown above, a domination query can be answered 
by answering $O(\tau \log \log n)$ counting queries; hence, the query time 
is $O(\tau \log \log n \log \log U)$. Since it is not necessary to store 
range counting data structures for each interval, all range counting data
 structures use $O(n)$ space.
\begin{theorem}\label{theor:colgrid}
There exists a $O(\tau n)$ space data structure  that 
supports one-dimensional   domination  queries in
 $O(\tau \log \log U \log \log n)$ 
time. 
\end{theorem}





\section{Dynamic Domination Queries in One Dimension}
\label{sec:dynamic}
Let $T$ be  a binary tree on the set of all $p\in P$. 
With  every internal node $v$ we associate a range $rng(v)=\halfrightsect{l_v}{r_v}$, where $l_v$ is the leftmost leaf descendant of $v$ and $r_v$ is the 
leaf that follows the rightmost leaf descendant of $v$. 
$T$ is implemented as a balanced binary tree, so that 
insertions and deletions are supported in $O(\log n)$ time and the tree height is $O(\log n)$. 
In each node $v$ we store the number of its leaf descendants, and 
the list $L_v$; $L_v$ contains all colors that dominate $rng(v)$. 
For every color $c$ in $L_v$ we also maintain the number 
of points of color $c$ that belong to $rng(v)$. 
For each color $c$ there is also a data structure that stores all points
of color $c$ and supports one-dimensional range counting queries.

A query $Q=[a,b]$ is answered 
by  traversing the paths from $l_a$ to $q$ and from $l_b$ to $q$,
where $l_a$ and $l_b$ are the leaves that contain $a$ and $b$ 
respectively, and $q$ is the lowest common ancestor of $a$ and $b$. 
As in the previous section, in every visited node $u$ 
the search procedure constructs the set of colors $S_u$, such 
that every $c\in S_u$ dominates $rng(v)\cap [a,b]$. 
Suppose that a node $v$ on the path from $l_b$ to $q$ is visited 
and let $u$ be the child of $v$ that is also on the path from $l_b$ to $q$.
If $u$ is the left child of $v$, then $rng(v)\cap [a,b]= rng(u)\cap [a,b]$ 
and $S_v=S_u$.
If $u$ is the right child of $v$, then 
$rng(v)\cap [a,b] =rng(w) \cup (rng(u)\cap [a,b])$ where $w$ is a sibling 
of $u$. Colors that dominate $rng(w)$ are stored in $L_w$; we know colors
that dominate $(rng(u)\cap [a,b])$ because $u$ was visited before $v$ 
and $S_u$ is already constructed. Hence, we can construct $S_v$ by 
examining each color $c\in L_w\cup S_u$ and answering the counting 
query for each color. Since one-dimensional dynamic range counting can be 
answered in $O(\log n)$ time, we spend $O(\tau \log n)$ time in each tree 
node. 
Nodes on the path from $l_a$ to $q$ are processed in a symmetric way. 
Finally we examine the colors stored in $S_{q_1}$ and $S_{q_2}$, where 
$q_1$ and $q_2$ are the children of $q$, and find the colors 
that dominate $rng(q)\cap [a,b]= [a,b]$.


When a new element is inserted(deleted), we insert a new leaf $l$ into $T$
(remove $l$ from $T$). 
For every ancestor $v$ of $l$, the  list  $L_v$ is updated.
 
After a new point of the color $c_p$ is inserted, the  color $c_p$ may 
dominate 
$rng(v)$ and colors in $L_v$ may cease to dominate 
$rng(v)$. We may check whether $c_p$ must be inserted into $L_v$ and whether 
some colors $c\in L_v$ must be removed from $L_v$ by performing 
at most $\tau +1$ range counting queries. Since a new point has 
$O(\log n)$ ancestors, insertions are supported in $O(\tau \log^2 n)$ time.
 
When a point of color $c_p$ is deleted, we may have to delete 
the color $c_p$ from $L_v$. We can test this by performing one 
counting query. However, we may also have to insert some new color 
$c$ into $L_v$ because the number of points stored in descendants of 
the node $v$ decreased by one. 
To implement this, we store the set of candidate colors $L'_v$; 
$L'_v$ contains all colors that $(\tau/2)$-dominate $rng(v)$. 
For each color $c\in L'_v$ we test whether $c$ became a $\tau$-dominating 
color after deletion.  When the number of leaf descendants of the node 
$v$ decreased by a factor 2, we re-build the list $L'_v$. If $P_v$ 
is the set of leaf descendants of $v$ (that is, points that belong to $rng(v)$), then we can construct the set of distinct colors that occur in $P_v$ 
in $O(|P_v|\log(|P_v|))$ time. We can also find the sets of colors that 
$\tau$-dominate and $(\tau/2)$-dominate $rng(v)$ in  
$O(|P_v|\log(|P_v|))$ time. Since we re-build $L'_v$ after a sequence 
of at least $|P_v/2|$ deletions, re-build of some $L'_v$ incurs an
 amortized cost $O(\log n)$. Every deletions may affect $O(\log n)$ 
ancestors; hence, deletions are supported in $O(\log^2 n)$ amortized time. 

We can speed-up the update operations by storing in each tree node 
$u$ the set of distinct colors in $P_u$, denoted by $C_u$. 
For each color $c\in C_u$, we store how many times points with color $c$ occur in $P_u$. When a new point $p$ is inserted/deleted, we can update 
$C_v$ for each ancestor $v$ of $p$ in $O(1)$ time.
Using $C_v$, we can decide whether a given  new color must be inserted 
into $L_v$ in $O(1)$ time. Using $C_v$ we can also re-build $L'_v$ 
in $O(|C_v|)=O(|P_v|)$ time. 
Hence, we can support insertions in $O(\tau \log n)$ time and deletions in 
$O(\log n)$ time with help of lists $C_v$. The total number of 
elements in all $C_v$ is $O(\tau n\log n)$. 

Thus we obtain the following 
\begin{theorem}
There exists a $O(\tau n)$ space data structure 
that supports one-dimensional domination queries 
 and insertions in $O(\tau \log^2 n)$ time  and deletions in 
$O(\tau \log^2 n)$ amortized time. 
 There exists a $O(\tau n\log n)$ space data structure 
that supports one-dimensional domination queries 
 and insertions in $O(\tau \log n)$ time  and deletions in 
$O(\tau \log n)$ amortized time.  
\end{theorem}

\section{Multi-Dimensional Domination Queries}
\label{sec:multidim}
We can extend our data structures to support 
$d$-dimensional queries for an arbitrary constant $d$  using the standard 
range trees~\cite{bentley} approach. 
\begin{theorem}
There exists a $O(n\log^{d} n)$ space data structure  that 
supports $d$-dimensional orthogonal range domination  queries 
in $O(\log^{d-1} n (\log \log n)^2)$ time.
\end{theorem}
We describe how we can construct a $d$-dimensional data 
structure if we know how to construct a  $(d-1)$-dimensional data structure.
A range tree $T_d$ is constructed  on the set of 
$d$-th coordinates of all points. An arbitrary interval 
$[a_d,b_d]$ can be represented as a union of $O(\log n)$ 
node ranges. Hence, an arbitrary $d$-dimensional query 
$Q=Q^{d-1}\times [a_d,b_d]$ can be represented as a union 
of $O(\log n)$ queries $Q_1,\ldots, Q_t$, where $t=O(\log n)$ 
and $Q_i= Q_{d-1}\times rng(v_i)$ for some node $v_i$ of $T$. 
In each node $v$ of $T$ we store a $(d-1)$-dimensional data structure $D_v$
that contains the first $d-1$ coordinates of all points whose 
$d$-th coordinates belong to $rng(v)$. 
$D_v$ supports modified domination queries in $d-1$ dimensions:
for a $(d-1)$-dimensional query rectangle $Q$, $D_v$ outputs all colors 
that dominate $Q\times rng(v)$.
Using $D_{v_i}$ we can find (at most $\tau$) colors that 
dominate $Q_i=Q'\times rng(v)$. 
Since $Q$ is a union of $O(\log n)$ ranges $Q_i$, we can identify 
a set $\cC$ that contains  $O(\tau\log n)$ candidate colors 
by answering $O(\log n)$ modified $(d-1)$-dimensional domination queries.
As follows from Observation~\ref{obs:decompose}, only a color from $\cC$ 
can dominate $Q$. Hence, we can identify all colors that 
$\tau$-dominate $Q$ by answering $O(\tau \log n)$ $d$-dimensional 
range counting queries. 
Thus the query time for $d$-dimensional queries can be computed 
with the formula $q(n,d)=O(\log n) q(n,d-1) + O(\tau \log n)c(n,d-1)$,
where $q(n,d)$ is the query time for $d$-dimensional domination queries 
and $c(n,d)$ is the query time for $d$-dimensional counting 
queries.
We can answer  $d$-dimensional range counting queries in 
$O(\log^{d-1} n)$ time and $O(n\log^{d-1} n)$ space ~\cite{chazelle}.
We can answer  one-dimensional domination queries 
in $O(\log n)$ time by Theorem~\ref{theor:static1d}. 
Therefore $d$-dimensional domination queries can answered 
in $O(\tau\log^d n )$ time.

We can apply the reduction to rank space technique~\cite{O88},~\cite{chazelle}
 and replace all point coordinates with  labels from $[1,n]$. 
This will increase the query time by an additive term $O(\log n)$. 
Since  point coordinates are bounded by $n$, we can apply Theorem~\ref{theor:colgrid}
and answer one-dimensional domination queries in $O((\log \log n)^2)$ time
using a $O(n)$ space data structure. 
Since the space usage grows by a $O(\log n)$ factor with each dimension,
our data structure uses $O(n \log^{d-1} n)$ space. 
\begin{theorem}
There exists a data structure that supports domination queries 
in $d$ dimensions in $O(\tau \log^{d} n)$ time 
and uses $O(n\log^{d-1} n)$ space. 
\end{theorem}
\tolerance=1000
The same range trees approach can be also applied to the dynamic 
one-dimensional data structure for domination queries. 
Since one-dimensional dynamic domination queries can be 
answered in $O(\tau \log^2 n)$ time  
and dynamic range counting queries 
can be answered in $O(\log^d n)$ time and $O(n\log^{d-1} n)$ space, 
$d$-dimensional domination queries can be answered in $O(\log^{d+1} n)$ 
time, and the space usage is $O(\tau n\log^{d-1} n)$. 
Since updates are supported in $O(\log^2 n)$ (amortized) time in 
one-dimensional case 
and update times grow by $O(\log n)$ factor with each dimension, 
$d$-dimensional data structure supports updates in $O(\log^{d+1} n)$ 
(amortized) time. 
\begin{theorem}
There is a dynamic data structure that answers 
domination queries in $d$ dimensions in $O(\tau \log^{d+1} n)$ 
time, uses  $O(\tau n\log^{d-1} n)$ space, and supports 
insertions in $O( \tau \log^{d+1} n)$ time and deletions in
 $O(\tau \log^{d+1} n )$ amortized time.
\end{theorem}
\section*{Conclusion}
We presented data structures for a new variant of colored range reporting 
problem. Our data structures use pseudo-linear space and report all  
$\tau$-dominating colors  in poly-logarithmic  time in the case when 
the parameter $\tau$ is small, \ie\  constant or poly-logarithmic in $n$. 
It would be interesting to construct efficient data structures for 
larger values of $\tau$.

Another interesting problem is construction of an efficient data 
structure that finds for an arbitrary given rectangle $Q$ and a (fixed) 
parameter $p$, the $p$ most frequently occurring colors in the rectangle 
$Q$. That is, the data structure must find the set of 
colors $C_p=\{\,c_1,\ldots,c_p\,\}$, such that 
$|\{\,p\in P\cap Q \mid \col(p)=c_i,\,c_i\in C_p\,\}| \geq 
|\{\,p\in P\cap Q \mid \col(p)=c,\,c\not\in C_p\,\}|$
\section*{Acknowledgment}
We would like to thank Mark~de~Berg and Herman Haverkort for  stimulating discussions and for suggestions concerning  the  new variant of colored range 
searching  problem.

\end{document}